\documentclass[aps,pra,reprint,showpacs,superscriptaddress]{revtex4-1}
\usepackage[english]{babel}
\usepackage{amsmath,amssymb,bbm,graphicx,color,comment,txfonts}
\usepackage[bookmarks=true,colorlinks,citecolor=blue,urlcolor=blue]{hyperref}
\usepackage{dsfont}
\usepackage[normalem]{ulem}

\newcommand{\cre}[1]{c^{\dagger}_{#1}}
\newcommand{\ann}[1]{c^{\phantom{\dagger}}_{#1}}
\newcommand{\EE}[1]{\overline{S_{\text{vN}}}(#1)}

\newcommand{\eq}[2]{\begin{eqnarray}\label{#1} #2 \end{eqnarray}}

\begin{document}

\date{\today}

\title{Entanglement transition in a monitored free fermion chain -- from extended criticality to area law}
\author{O. Alberton}
\affiliation{Institut f\"ur Theoretische Physik, Universit\"at zu K\"oln, D-50937 Cologne, Germany}
\author{M. Buchhold}
\affiliation{Institut f\"ur Theoretische Physik, Universit\"at zu K\"oln, D-50937 Cologne, Germany}
\author{S. Diehl}
\affiliation{Institut f\"ur Theoretische Physik, Universit\"at zu K\"oln, D-50937 Cologne, Germany}

\begin{abstract}
We analyze the quantum trajectory dynamics of free fermions subject to continuous monitoring. For weak monitoring, we identify a novel dynamical regime of subextensive entanglement growth, reminiscent of a critical phase with an emergent conformal invariance. For strong monitoring, however, the dynamics favors a transition into a quantum Zeno-like area-law regime. Close to the critical point, we observe logarithmic finite size corrections, indicating a Berezinskii–Kosterlitz–Thouless mechanism underlying the transition. This uncovers an unconventional entanglement transition in an elementary, physically realistic model for weak continuous measurements. In addition, we demonstrate that the measurement aspect in the dynamics is crucial for whether or not a phase transition takes place. 
\end{abstract}

\maketitle
{\it Introduction.}~-- Fingerprints of the competition between unitary and non-unitary dynamics are found in almost all aspects of modern quantum science. The spectrum ranges from radiative decay in driven two-level systems~\cite{Mollow,Dicke} to dephasing of trapped ions and cold atoms due to laser noise~\cite{RevModPhys.75.281} or phonon-induced dissipation in electronic devices and color centers~\cite{PhysRevLett.93.016601,vacancy2017}. Non-unitary processes crucially affect quantum dynamics from single particles to the many-body realm.

One fascinating example are phase transitions in the entanglement entropy, which have been discovered in unitary circuit dynamics subject to local projective measurements~\cite{Nahum2017,Skinner2019,Fisher2018, Chan2019,choi2019,Jian2020}. Focussing on the entanglement properties of individual measurement trajectories $|\psi(\xi)\rangle$, where $\xi(t)$ is a realization of temporal randomness encountered in quantum mechanical measurements, a transition from an entangling evolution obeying a volume-law to a disentangled evolution governed by an area-law as a function of the measurement rate has been identified~\cite{gullans2019,Zabalo2020,zhang2020,Tang2020,Bao_2020,Li2019b,gullans2019scalable,Nahum20a}. A characteristic trait of these transitions is that they manifest themselves in \emph{state-dependent} observables $\hat{O}(\rho(\xi))$, with $\rho (\xi) = |\psi(\xi)\rangle\langle \psi(\xi)|$. For example, for the entanglement entropy of a subsystem $A$, $\hat{O}(\rho(\xi))=-\log\rho_A(\xi)$, where $\rho_A(\xi)$ is the reduced density matrix on $A$ -- a highly nonlinear function of the state $\rho(\xi)$. Such entanglement transitions have been reported in several setups, including non-unitary circuit models and chains of interacting bosons subject to continuous measurements~\cite{Schomerus2019,goto2020,li2020conformal, Alba2020, fuji2020,chen2020,Romito2020}.

Here we focus on one of the most elementary models for the competition between unitary and non-unitary dynamics, free fermions on a periodic chain, subject to coherent hopping and local monitoring of the fermion particle number, which preserves the system's $U(1)$-symmetry~\cite{Cao2019,Knap2018,PhysRevLett.120.133601,Yang2018}. This model can be simulated efficiently on large system sizes \cite{Cao2019}. Moreover, it is natural in terms of physical implementations (although this does not guarantee straightforward observability of entanglement transitions): this scenario arises, e.g., for ultracold fermions in optical lattices, Rydberg atom arrays or spin chains, where the local particle number (or magnetization) is measured via homodyne detection ~\cite{PhysRevLett.120.133601,Yang2018}. From a measurement theory point of view, the non-unitary monitoring evolution results from taking the temporal continuum limit of weak measurements of the local fermion particle number, implemented for instance by a weak, local coupling to a projectively measured photon bath~\cite{Yang2018,deVega2017,Milburn1993,Molmer1992}.

\begin{figure}
  \includegraphics[width=0.96\linewidth]{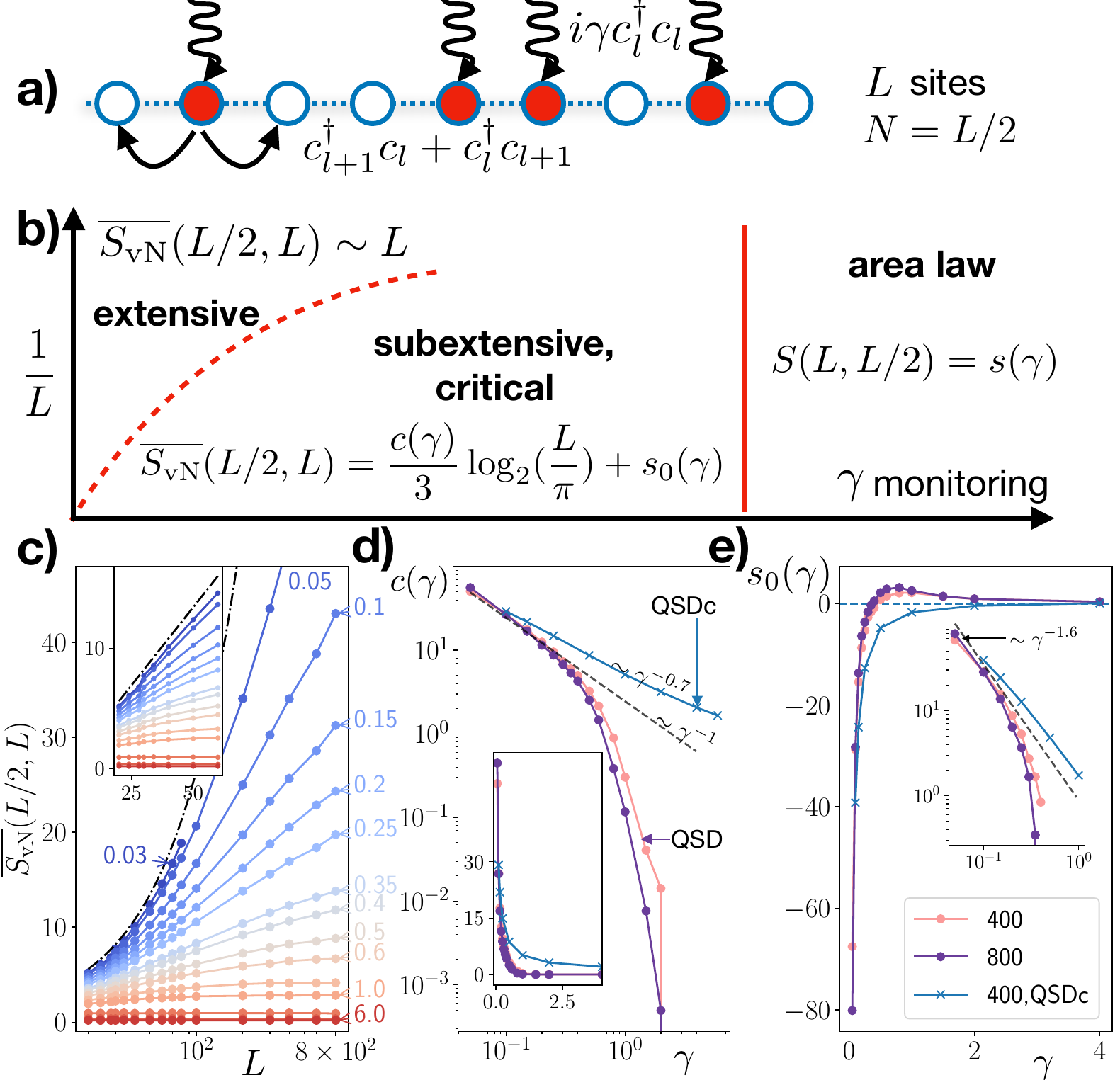}
  \caption{(a) Free fermions hopping on a chain of length $L$ subject to continuous monitoring with dimensionless rate $\gamma$. (b) Schematic ``phase diagram'' showing the different regimes of entanglement scaling with $L$ (the dashed line denotes a finite-size crossover). (c) At small monitoring rate, a subextensive growth of the entropy $\sim\log(L)$ at sufficiently large $L$ is reminiscent of a critical, conformally invariant phase. For small $\gamma, L$, extensive growth $\sim L$ is observed (inset), approaching a volume law as $\gamma\rightarrow0$. (d,e) The effective central charge and residual entropy obtained by fitting the data to Eq.~\eqref{Eq8}. The blue lines in (d,e) correspond to the non-unitary circuit evolution (QSDc), for which the transition is absent. The insets show the same data on a linear (d) and logarithmic (e) scale~\cite{Sim}.
  }
  \label{Fig1}
\end{figure}%%%

We report two central findings. (i) We establish the existence of an extended, robust 'weak-monitoring' regime, for which the entanglement entropy asymptotically grows logarithmically with the subsystem size. This previously unanticipated regime is reminiscent of a critical, conformally invariant phase of fermions in ($1+1$) dimensions.  We strengthen the analogy to conformal field theory (CFT) by examining the behavior of connected density-density correlations and the mutual information, both displaying clear signatures of conformal invariance. 

(ii) For strong monitoring, the system undergoes a phase transition into an area law phase obeying disentangling dynamics. At the critical point, which is located at a non-zero measurement strength, finite size scaling of the entanglement entropy provides strong indications for a Berezinskii–Kosterlitz–Thouless (BKT) scenario underlying the measurement induced phase transition. We characterize both phases in terms of the entanglement entropy, the mutual information and the connected density correlation function, and find evidence that  the conformally invariant, weak measurement phase is left via the BKT mechanism.

Finally, we compare the measurement induced dynamics with a non-unitary circuit evolution, which neglects the measurement back-action and violates probability conservation during the dynamics. While exact probability conservation is guaranteed for any true monitoring dynamics, the average conservation is sufficient to guarantee a consistent open system quantum dynamics. In particular, all considered protocols collapse onto the same Lindblad quantum master equation. The difference surfaces however once state dependent observables are considered: Amongst the trajectory evolutions considered here, only physical measurement protocols exhibit an entanglement phase transition.

The entanglement phase diagram is displayed in Fig.~\ref{Fig1}. We confirm that the volume law realized at $\gamma =0$ is unstable against infinitesimal monitoring $\gamma>0$ \cite{chen2020, Cao2019}. It is instead replaced by an intriguing subextensive behavior, which has also been observed very recently for free fermions with complete spatio-temporal randomness \cite{chen2020}, and in measurement-only protocols \cite{ippoliti2020}. The opposite limit $\gamma^{-1} = 0$ lacks any entangling operations and is characterized by an area law. We find a phase transition from the logarithmic scaling behavior, with a $\gamma$-dependent {\it effective} central charge, to an area law at a finite $\gamma_c$ (cf. Fig. \ref{Fig1}), similarly to \cite{ippoliti2020}, accompanied by a sudden drop of the central charge to zero. The transition is, however, absent for a non-unitary circuit protocol, where only the logarithmic regime is observed.

{\it Trajectory evolution.}~-- We consider free fermions on a half-filled periodic chain of length $L$, which is described by the nearest-neighbor hopping Hamiltonian $H=\sum_{l} \cre{l+1}\ann{l}+\cre{l}\ann{l+1}$ with fermionic creation and annihilation operators $\cre{l}, \ann{l}$. Furthermore, the local fermion densities $n_l$ are continuously, weakly measured, i.e. monitored, by some external mechanism, yielding a non-unitary contribution to the Hamiltonian. Generically this includes a stochastic term $\sim i \xi_{l,t} n_l$ with random events $\{\xi_{l,t}\}$, such that the time-evolution of a fermion pure state $|\psi (\{\xi_{l,t}\})\rangle$ follows a stochastic trajectory.

For the major part of our analysis, we consider the quantum state diffusion (QSD) protocol. Here, the monitoring of the fermion densities is implemented via their coupling to a set of continuous variable bath operators~\cite{Gisin_1992,Strunz1998}. Paradigmatic examples include the positions of free particles (so-called pointers)~\cite{Schomerus2019, Carlton1987} or the quadratures of a photon environment, which can be measured via homodyne detection to implement a QSD evolution with cold atoms~\cite{PhysRevLett.120.133601,Yang2018}. 
The wave functions in the QSD protocol follow the evolution equation~\cite{Gisin_1992,Strunz1998,deVega2017}
\begin{equation}
  d| \psi\{\xi_{l,t}\} \rangle=\Big[ -iHdt + \sum_l \Big(\xi_{l,t} \hat{M}_{l,t}-\frac{\gamma}{2} \hat{M}_{l,t}^2
      dt \Big) \Big] |\psi \{\xi_{l,t}\}\rangle,
  \label{Eq2}
\end{equation}
where $\hat{M}_{l,t}=n_l-\langle n_l\rangle_t$~\footnote{The QSDc evolution is realized by setting $\hat M_{l,t}=n_l$.}.
The  real-valued Gaussian noise $\xi_{l,t}$ has zero mean $\overline{\xi_{l,t}}=0$ and covariance $\overline{\xi_{l,t} \xi_{m,t'}}=\gamma dt\delta_{l,m}\delta(t-t')$.

For reference, we compare our results to two additional trajectory evolution protocols: (i) The quantum jump evolution (QJ), which realizes a monitoring dynamics with a discrete measurement noise ~\cite{Molmer1992,Milburn1993,Daley_2014} and displays qualitatively similar behavior as the QSD~\cite{supp}. (ii) The continuous-time limit of a non-unitary circuit description (QSDc)~\cite{supp, chen2020}, also known as "raw" quantum state diffusion~\cite{Strunz1998,deVega2017,supp}, which does not correspond to any monitoring. 

{\it Numerical procedure.}~-- The evolution equation~\eqref{Eq2} is quadratic in the fermion operators, thus any initial Gaussian state $|\psi_0\rangle$ remains Gaussian under time evolution. This enables efficient numerical simulation of Eq.~\eqref{Eq2}, which is outlined in Refs.~\cite{Cao2019, supp}. 
The full information of the Gaussian fermion density matrix and correlations is
encoded in the correlation matrix ${D_{l,j}(t,t')=\langle
  \cre{l,t}\ann{j,t'}\rangle}$. For a  chain of length $L$, the von Neumann entanglement entropy
$S_{\text{vN}}(l,L)$ for a subsystem $A$ of length $l$ can
be obtained from the eigenvalues of the equal-time correlation matrix of
subsystem $A$ \cite{Calabrese_2005,Alba2018} (see also~\cite{supp}).

In what follows we initialize the system in a short range correlated N\'{e}el
state $|\psi_0\rangle=|010101....01\rangle$, and evolve the different types of
trajectories according Eq.~\eqref{Eq2}. The entanglement entropy,
mutual information and correlation functions are computed for each individual
trajectory after the evolution has reached a steady state, $\gamma
t\gg1$~\cite{supp}. We denote the trajectory average of an observable $O$ by
$\overline{O}$. The linear average $\overline{D}=\frac{1}{2}\mathds{1}$ corresponds to an infinite temperature state for any $\gamma>0$ and is independent of the trajectory evolution. For a nonlinear function of the correlation matrix $f(D)$, however, 
generally $\overline{ f(D)}\neq f(\overline{ D})$ and therefore $\overline{
  S_{\text{vN}}}(l,L)$
cannot be obtained from the linear average $\overline{D}$.

{\it Entanglement phase transition.}~-- For a bipartition of the chain into two equal subsystems, the steady-state entanglement entropy $\EE{L/2,L}$ shows three different functional dependencies on the chain length $L$ and the monitoring rate $\gamma$, as illustrated in Fig.~\ref{Fig1}(c) (see \cite{supp} for QJ). For the coherent time evolution at $\gamma=0$, an initial N\'{e}el state develops an extensive entanglement entropy converging to a volume law \cite{Alba2018}. This behavior transcends as a finite size effect to weak, but non-zero monitoring, where an extensive entanglement growth $\overline{S_{\text{vN}}}(L/2, L)\sim L$ is observed for $L<L_c(\gamma)\sim \exp(\sqrt{\gamma_0/\gamma})$ smaller than a $\gamma$-dependent cutoff length (see, e.g., inset of Fig.~\ref{Fig1}(c)).

For any non-zero monitoring rate $0<\gamma<\gamma_c$, and below a critical rate $\gamma_c$, the entanglement in the thermodynamic limit follows a subextensive growth $S\sim \log L$. This is characteristic for $(1+1)$-dimensional CFTs~\cite{Calabrese_2004,Calabrese_2009}. Here, we describe this growth according to a CFT with periodic boundaries
\begin{equation}
\label{Eq8}
  \EE{l,L}=\tfrac{c(\gamma)}{3}\log_2 \left[\tfrac{L}{\pi}\sin\left(\tfrac{\pi l}{L}\right)\right]+s_0(\gamma),
\end{equation}
but with a $\gamma$-dependent {\it effective} central charge $c(\gamma)$ and residual entropy $s_0(\gamma)$, see Fig.~\ref{Fig1}~(d,e).  Irrational and partly continuous central charges are established in CFTs for disordered or percolation problems~\cite{Cardy1997,Refael_2004,Laflorencie2005,Cardy2000,Skinner2019} and have been recently reported for entanglement transitions as well~\cite{fuji2020,chen2020,Skinner2019,Li2019b,ippoliti2020}. An extended regime of logarithmic scaling of $\EE{l,L}$ was observed recently for a free, non-unitary circuit dynamics with spatio-temporal randomness~\cite{li2020conformal}. Here,
we establish an extended phase of measurement-induced conformal invariance for free fermions on a regular lattice.

\begin{figure}
  \includegraphics[width=\linewidth]{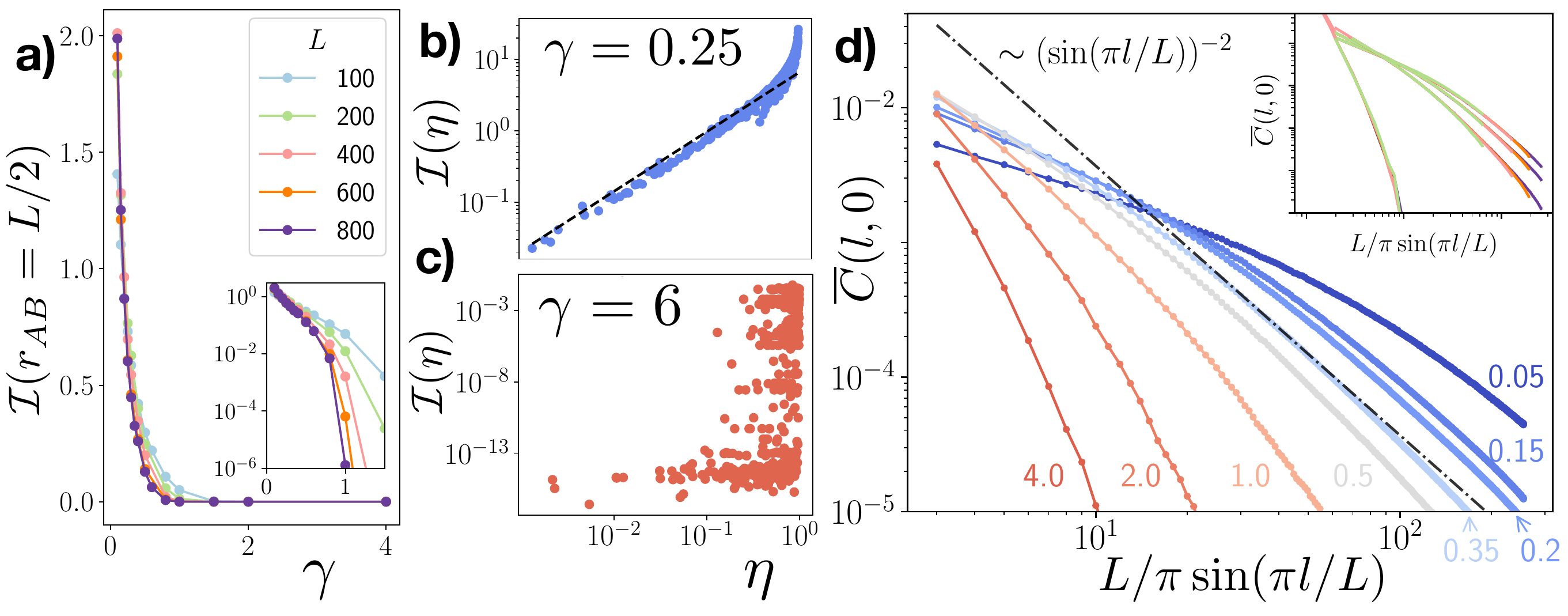}
  \caption{The conformal invariance at weak monitoring is confirmed (a) by a large, non-zero mutual information $l_{\text{A}}=l_{\text{B}}=L/2$, which rapidly decays to zero in the area law regime, and (b) by a scaling collapse of the mutual information as a function of the cross ratio $\eta$, i.e. $\mathcal{I}(\eta)\sim\eta$ ($L=400$). (c) In the area law regime no collapse is observed. (d) Equal time correlations $\overline{C}(l,0)$ decay algebraically $\sim l^{-2}$ (exponentially) with the distance $l$ in the conformally invariant (area law) regime ($L=800$). The inset shows a data collapse for different system sizes $L=200,400,600,800$ (axes range as in main plot).
  }
  \label{Fig3}
\end{figure}

\begin{figure}
  \includegraphics[width=\linewidth]{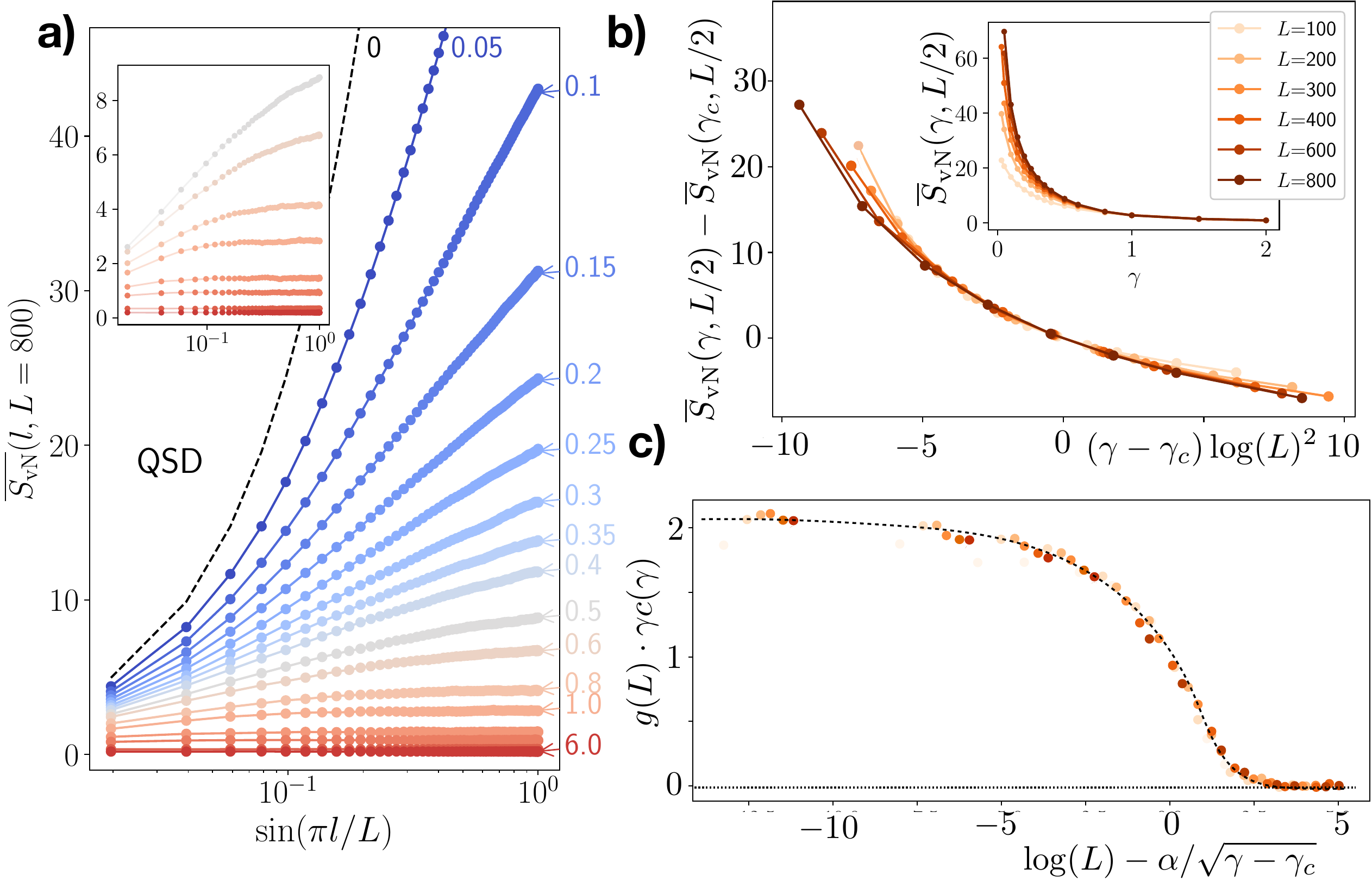}
  \caption{(a) The entanglement entropy as a function of the bipartition size $l$ reveals a clear, asymptotic logarithmic growth for weak monitoring and shows a transition to an area-law for stronger monitoring $\gamma\ge\gamma_c$ (inset). (b) Finite size scaling collapse of the entanglement entropy, assuming BKT scaling of the correlation length and $\gamma_c=0.31$. (c) Finite size scaling of the effective central charge, predicting a jump of $c(\gamma)$ from $c(\gamma_c-0^+)\approx \frac{2}{\gamma_c}$ to $c(\gamma_c+0^+)=0$ in the limit $L\rightarrow\infty$ (dotted lines are guides to the eye). The parameters are $\alpha=3.99, g(L)=(1+\frac{1}{2\log(L)-4.37})^{-1}$ and the legend from (b) applies in (c). }
  \label{Fig2}
\end{figure}

A major finding of our work is the existence of a phase transition at a critical monitoring rate $\gamma_c$, above which conformal invariance is lost and the entanglement entropy obeys an area law. This transition is well illustrated in the behavior of the effective central charge $c(\gamma)$.  For weak monitoring, $c(\gamma)\sim \gamma^{-1}$ decays algebraically and saturates at a non-zero value for $L\rightarrow\infty$.  At stronger monitoring, the effective central charge, and therefore the logarithmic scaling, vanish above a critical value $\gamma_c$ in the limit $L\rightarrow \infty$. For any finite size $L<\infty$, $c(\gamma)$ approaches zero according to an exponential $\log c(\gamma)\sim -|\gamma-\gamma_c(L)|^{-\alpha(L)}$ for some $\alpha(L)>0$ (see Fig.~\ref{Fig2}(c)). The phase transition is evidenced by a set of different, unambiguous observations (i) a qualitative change in the entanglement entropy, which no longer shows any subsystem-dependence for $\gamma\ge \gamma_c(L)$ ($\gamma_c(L=800)\approx 0.8$ in Fig.~\ref{Fig2}(a)), (ii) the behavior of the effective central charge with $\gamma$ in Fig.~\ref{Fig1}(d), as well as with the system size $L$ in Fig.~\ref{Fig2}(c), which drops to zero for $\gamma>\gamma_c$ and $L\rightarrow\infty$, (iii) the zero-crossing of the residual entropy $s_0(\gamma)$ in Fig.~\ref{Fig1}(e), which is required for a well-defined, positive entanglement entropy when $c\rightarrow0$, and (iv) qualitative changes of the mutual information and the correlation function, shown in Fig.~\ref{Fig3} and discussed right below.

{\it Mutual information.}~-- The mutual information
$\mathcal{I}(l_{\text{A}},l_\text{B})$ between two disjoint subsystems $A=[m_1,
m_2]$, $B=[m_3,m_4]$ of length $l_{\text{A}},l_\text{B}$ has emerged as a useful
indicator to locate an entanglement transition \cite{Li2019b}. It is
given by
$\mathcal{I}(l_{\text{A}},l_\text{B})=S_{\text{vN}}(l_{\text{A}},L)+S_{\text{vN}}(l_{\text{B}},L)-S_{\text{vN}}(A
\cup B,L)$, where $S_{\text{vN}}(A
\cup B,L)$ is the entanglement entropy of the subsystem $A\cup B$. $\mathcal{I}(l_{\text{A}},l_\text{B})$ measures
the amount of information that can be gained about subsystem $A$ from subsystem $B$ and
vice versa, and it is also an upper bound for connected correlation functions
between $A$ and $B$ \cite{wolf2008area}. For two disjoint intervals $l_{\text{A}}=l_{\text{B}}=L/8$, with centers at a distance $r_{AB}= L/2$, it is expected to show a sharp peak at the critical point separating the area and the volume law phase \cite{Li2019b}. Inspecting $\mathcal{I}(l_A=l_B=L/8, r_{AB}=L/2)$ for different system sizes in Fig.~\ref{Fig3}(a) shows that it is significantly larger than zero in the entire critical regime and approaches zero rapidly in the area law phase, reflecting extended criticality. A similar peak is observed for the QJ evolution \cite{supp}.

%%%

For variable subsystem sizes, it is useful to define the cross ratio $\eta=\frac{m_{12}m_{34}}{m_{13}m_{24}}$ with
$m_{\alpha\beta}=\sin\left(\pi|m_\alpha-m_\beta|/L\right)$.
In the conformally invariant regime, the mutual information $\mathcal{I}(\eta)$ collapses onto a single line for all $\eta$, with a linear increase $\sim \eta$ for small cross ratios, see Fig.~\ref{Fig3}(b). The linear dependence in $\eta$ also implies a power-law decay of the mutual-information $\mathcal{I} \sim r_{AB}^{-2}$ for small subsystems with large separation \cite{Li2019b}. This collapse is a strong signature of conformal invariance and can be observed throughout the entire logarithmic regime. It can be contrasted with the behavior in the area law phase, shown in Fig.~\ref{Fig3}(c), where no collapse is observed. 

{\it Correlation function.}~-- In addition, we detect signatures of conformal invariance in connected correlation functions
\begin{equation}
  \label{Eq7} 
  C(l,\tau)  \equiv |D_{l+j,j}(t+\tau,t)|^2=\langle n_{l+j,t+\tau}\rangle\langle n_{j,t}\rangle-\langle n_{l+j,t+\tau}n_{j,t}\rangle, \nonumber
\end{equation}
which is the Fock (exchange) contribution to the density-density correlation in a Gaussian state. $C(l,\tau)$ is a second moment of the correlation matrix $D$, and thus its trajectory average does not correspond to an infinite temperature state.

The equal-time correlation functions $\overline{C}(l,0)$ in Fig.~\ref{Fig3}(d) quantitatively reflect the phase diagram in Fig.~\ref{Fig1}(b). In the conformally invariant regime, an algebraic decay of the correlation function with the square of the distance $\sim [\sin(\pi l/L)]^{-2}$ is observed. The collapse of the correlation functions for variable system sizes in the inset of Fig.~\ref{Fig3}(d) demonstrates that this $\sim [\sin(\pi l/L)]^{-2}$ scaling is observed in the thermodynamic limit $L\rightarrow\infty$. On distances $l<L_c(\gamma)$ (for volume law) or in the area law regime the correlations deviate significantly from the $\sim l^{-2}$ behavior, showing longer-ranged or short-ranged correlations, respectively.

%%%%
%%%%

{\it BKT transition and critical point}~--
In unitary quantum dynamics, the scenario of a phase transition from an extended conformally invariant phase to an area law phase via the generation of a scale in (1+1) dimensions is an unambiguous and exclusive feature of the BKT mechanism~\cite{PhysRevD.80.125005}. The measurement-induced phase transition reported here displays several similarities to this phenomenology, including the sudden drop of the effective central charge (and of the mutual information, Fig.~\ref{Fig3}(a) inset) and the loss of conformality, accompanied by the emergence of a length scale $\xi$ in the correlation functions $C(l,0)\sim \exp(-l/\xi)$  (Fig.~\ref{Fig3}(b-c)).

Inspired by this similarity, we perform a finite size scaling analysis of the entanglement entropy and the effective central charge $c(\gamma)$, for which we assume a BKT-type correlation length $\xi\sim\exp(-\alpha/\sqrt{|\gamma-\gamma_c|})$.
Here $|\gamma-\gamma_c|$ is the distance from the measurement-induced critical point. For the entanglement entropy, this yields the scaling form $S_{\text{vN}}(\gamma,L/2) -
S_{\text{vN}}(\gamma_c,L/2) = F[ (\gamma-\gamma_c) \log(L)^2 ]$~\cite{harada1997universal, carrasquilla2012superfluid} with a scaling function $F$. We observe a convincing collapse for a range of critical rates $\gamma_c$, with the best fit $\gamma_c=0.31$ being displayed in Fig.~\ref{Fig2}(b).

The central charge is expected to be zero for $\gamma>\gamma_c$ and to display a sudden jump at $\gamma=\gamma_c$ in the limit $L\rightarrow\infty$, analogous to the quantum phase transition in equilibrium. Observables undergoing such a sudden jump at the critical point are well described by a scaling function $\tilde F(X)$ with argument $X=\log(L)-\alpha/\sqrt{\gamma-\gamma_c}$~\cite{Sandvik_2010}. The scaling collapse is shown in Fig.~\ref{Fig2}(c) for the product $\gamma c(\gamma)$. It covers two limits: (i) the case $\gamma>\gamma_c$ and $L\rightarrow \infty$, corresponds to $X\rightarrow\infty$ and therefore $c(\gamma)=0$, (ii) the case $L<\infty$ and $\gamma\rightarrow\gamma_c$ corresponds to $X\rightarrow-\infty$ and roughly $\lim_{X\rightarrow-\infty}\gamma c(\gamma)\rightarrow2$, according to Fig.~\ref{Fig2}(c). Overall it predicts a jump of the effective central charge at $\gamma=\gamma_c$ and a critical value $c(\gamma_c)=\frac{2}{\gamma_c}$. 

The finite size collapse and the scaling behavior of the central charge work well for a range of critical couplings $\gamma_c\in [0.2 0.35]$ with the best results obtained for $\gamma_c=0.31$. Without additional analytical constraints (such as, e.g., an analogue of the Nelson-Kosterlitz criterion~\cite{Hsieh_2013}), the precise location of the critical point is, however, still hard to determine more accurately. This is a general problem of phase transitions with slowly diverging length scales. Nevertheless, Fig.~\ref{Fig2} provides strong indications for a phase transition of the BKT universality classe.
The observation of an entanglement transition from a logarithmic to area law regime modifies the conclusion of earlier work on free fermions, which ruled out a volume law phase under monitoring and concluded an area law for any $\gamma>0$ \cite{Cao2019}.

{\it Importance of true measurements.}~-- We compare three different evolution protocols, two of which correspond to a physical measurement dynamics (QSD and QJ) and one to a non-unitary circuit evolution (QSDc) without measurement back-action. All three yield qualitatively similar results in the conformally invariant regime, $\gamma\le\gamma_c$ (see Fig.\ref{Fig1}(d) for QSDc and \cite{supp} for QJ). However, only the physical measurement protocols exhibit a transition towards an area law phase at larger monitoring rates $\gamma\ge\gamma_c$. The QSDc indicates no area law transition, instead the conformal invariance is extended to arbitrary $\gamma>0$. This behavior roots in the absence of measurement dark states in the QSDc evolution. 

In the absence of the Hamiltonian, both QSD and QJ display a set of measurement dark states $\{|\psi_D\rangle\}$, i.e., eigenstates of all measurement operators $n_l|\psi_D\rangle=\lambda_l|\psi_D\rangle$. For instance in the QSD evolution $\hat M_{l,t}|\psi_D\rangle=0$, and therefore $|\psi_D\rangle$ is an attractor of the dynamics in the strong monitoring limit $\gamma\gg J$. This reflects the tendency of a repeatedly measured system to eventually collapse into eigenstates of the measured operators. For continuous measurements, this collapse requires a measurement back-action, which is absent in the QSDc evolution. In this case, higher moments of the correlation matrix $D$ will show a significant deviation from a physical measurement dynamics, for instance in the norm of the state and the entanglement entropy \cite{supp}. A significant difference arises also in the distribution function of higher moments. The distribution function for the trajectory entanglement entropy for instance undergoes a qualitative change at the phase transition in the QSD evolution, while it remains unmodified for the QSDc evolution \cite{supp}. 

{\it Discussion and conclusion.}~-- A natural model of continuously monitored, free fermions can realize an entanglement phase transition with strong indications of BKT universality. Instead of interpolating between volume- and area law behavior, the transition connects a 'gapless' phase with conformal invariance and a logarithmic scaling of the entanglement entropy, to an area law. Beyond exhibiting the characteristic phenomenology of entanglement transitions, it manifests in the behavior of connected correlations functions of the continuously monitored observables. We show that this entanglement transition also appears in a free fermion dynamics with spatio-temporal disorder~\cite{supp}, which demonstrates that the entanglement scenario drawn for this model is not peculiar to an integrable tight-binding Hamiltonian.

Our results open intriguing lines for future research: The simplicity of the model and the connections to CFT and the Kosterlitz-Thouless scenario of an extensive critical regime~\cite{PhysRevD.80.125005}, cut off at a critical monitoring rate $\gamma_c$, spark the hope to understand the transition and its variants, observed in unitary circuit models of free fermions~\cite{bao2021symmetry} and in a Dirac field theory~\cite{buchhold2021effective} more deeply, and to find a way towards experimental detectability.

\begin{acknowledgments}
  We acknowledge support from the Deutsche Forschungsgemeinschaft (DFG, German Research Foundation) under Germany's Excellence Strategy Cluster of Excellence Matter and Light for Quantum Computing (ML4Q) EXC 2004/1 390534769, and by the DFG Collaborative Research Center (CRC) 183 Project No. 277101999 - project B02. S.D. and O.A. acknowledge support by the European Research Council (ERC) under the Horizon 2020 research and innovation program, Grant Agreement No. 647434 (DOQS). M.B. acknowledges funding via grant DI 1745/2-1 under DFG SPP 1929 GiRyd. The code for our numerical computations was written in Julia~\cite{bezanson17}. We furthermore thank the Regional Computing Center of the University of Cologne (RRZK) for providing computing time on the DFG-funded High Performance Computing (HPC) system CHEOPS as well as support.
\end{acknowledgments}

\appendix

\section{Numerical implementation}
This part provides details for the numerical implementation of the trajectory
evolution used in the main text, Eqs. (1),(2). For each individual
trajectory, the state at time $t$ is a Gaussian state, which is parametrized by
an $L \times N$ matrix $U(t)$ via
\begin{equation}
  |\psi_t\rangle=\prod_{l=1}^{L/2} \Big(\sum_{j=1}^L
  U^{\phantom{*}}_{j,l}(t)\cre{j}\Big)|0\rangle ,
  \nonumber
\end{equation}
where $U^\dagger U = \mathbbm{1}$ since we explicitly normalize the state after each time step. In
other words, the state $| \psi_t \rangle$ is a Slater determinant state of $L/2$
fermions where the single-particle wavefunctions are given by the columns of
$U$. That is, the $l$-th occupied single-particle state is given by $| \phi_l(t)
\rangle = \sum_j U_{j,l}(t) | j \rangle$, where $| j \rangle$ is the 
wave-function localized on lattice site $j$.
% $\prod_l\alpha_{l,t}=\langle\psi_t|\psi_t\rangle$ is the norm of the state
% ($\alpha_{l,t}=1$ for QSD and QJ).
The correlation matrix is computed from $U$ via
\begin{equation}
  {D_{l,j}(t,t')=\big[U(t')U^\dagger(t)\big]_{j,l} =\langle \cre{l,t}\ann{j,t'}\rangle}. \nonumber
\end{equation}

Given the correlation matrix $D$ for a chain of length $L$, we compute the von Neumann
entanglement entropy $S_{\text{vN}}(l,L)$ for a subsystem A$=[m_1, m_2]$ of
length $l=|m_1-m_2|$  from the eigenvalues
$\{\lambda^{(\text{A})}_j\}$ of the reduced equal-time correlation matrix
$D^{(A)}(t,t) = D_{i=m_1,..,m_2, j=m_1,..,m_2}(t,t)$ on A via
\cite{Calabrese_2005,Alba2018}
\begin{equation}
  \label{Eq6} 
  S_{\text{vN}}(l,L)=-\sum_{j=1}^{l}\lambda^{(\text{A})}_j \log_2\lambda^{(\text{A})}_j+(1-\lambda^{(\text{A})}_j)\log_2(1-\lambda^{(\text{A})}_j).\nonumber
\end{equation}

To simulate the quantum state diffusion we follow the Trotterization approach used in Ref.~\cite{Cao2019}. Evolving $U$ over a time step $dt$ and neglecting corrections of order $(dt)^2$ yields up to an overall normalization
\begin{align}
  U(t+dt)=\text{diag}(e^{\xi_{1,t} + \gamma\sigma (2 \langle n_1 \rangle_t -1 ) dt},..\ ,\ e^{\xi_{N,t} + \gamma\sigma (2 \langle n_N \rangle_t -1 ) dt}) e^{-i h dt} U,\nonumber
\end{align}
where $h$ is the hopping matrix, $\sigma=1$ (QSD) or $\sigma=0$ (QSDc) and the $\langle  n_l \rangle_t$ are computed from
the correlation matrix $D_{l,l}(t,t)$. We then ensure that the columns of $U$ are orthonormal by performing a QR decomposition $U = QR$ and redefining $U=Q$. The applied step size is $dt=0.05$. 

To simulate the quantum-jump evolution Eq. (2), we exploit that particle number
conservation enforces a constant jump rate $\gamma N$ and apply the common jump
evolution procedure described in~\cite{Daley_2014}. (i) Determine the jump time $\tau= - \log(r)/(\gamma N)$ by drawing a random number $r$ uniformly from $[0,1]$. (ii) Evolve the time step $t$ to $t+\tau$ via $U(t+\tau)=e^{-i h \tau}U(t)$ and choose a jump operator $n_j$ according to the probabilities $P(n_j)=\langle  n_j\rangle_{t+\tau}/N$. (iii) Apply the jump to the correlation matrix $D= U(t+\tau)U^\dagger(t+\tau)$ according to
\begin{equation}
  D_{lm}  \to
  \begin{cases}
    1, & l=m=j \\
    0, & l\neq m \text{ and } (l=j \text{ or } m=j)\\
    D_{lm} - \frac{ D_{jm}D_{lj}}{\langle n_j \rangle_t}, & \text{otherwise}
  \end{cases}. \nonumber
\end{equation}
(iv) Obtain the new $U$ matrix by performing an SVD decomposition $D^* =USU^\dagger$ for a Hermitian matrix $D^*$ (note $S_{11}=..=S_{NN}= 1, S_{N+1,N+1} =..=S_{LL}=0$).

\section{Quantum jump results}
Here we provide results for the QJ evolution, which are discussed in the main
text but not displayed in the figures for clarity. 
In the quantum jump trajectories, the evolution equation is \begin{equation}\label{Eq3} d|\psi\{\xi_{l,t}\}\rangle =\Big[-iHdt+\sum_l \xi_{l,t} \Big(\frac{n_l}{\sqrt{\langle n_l\rangle_t}}-1\Big)\Big]|\psi\{\xi_{l,t}\}\rangle, \end{equation} for a state with conserved total particle number. For QJ the noise is defined via $\xi_{l,t}^2=\xi_{l,t}$ and $\overline{\xi_{l,t}}=\gamma dt \langle n_l\rangle_t$~\cite{Molmer1992,Milburn1993,Daley_2014}.

\begin{figure}
  \centering
  \includegraphics[width=\linewidth]{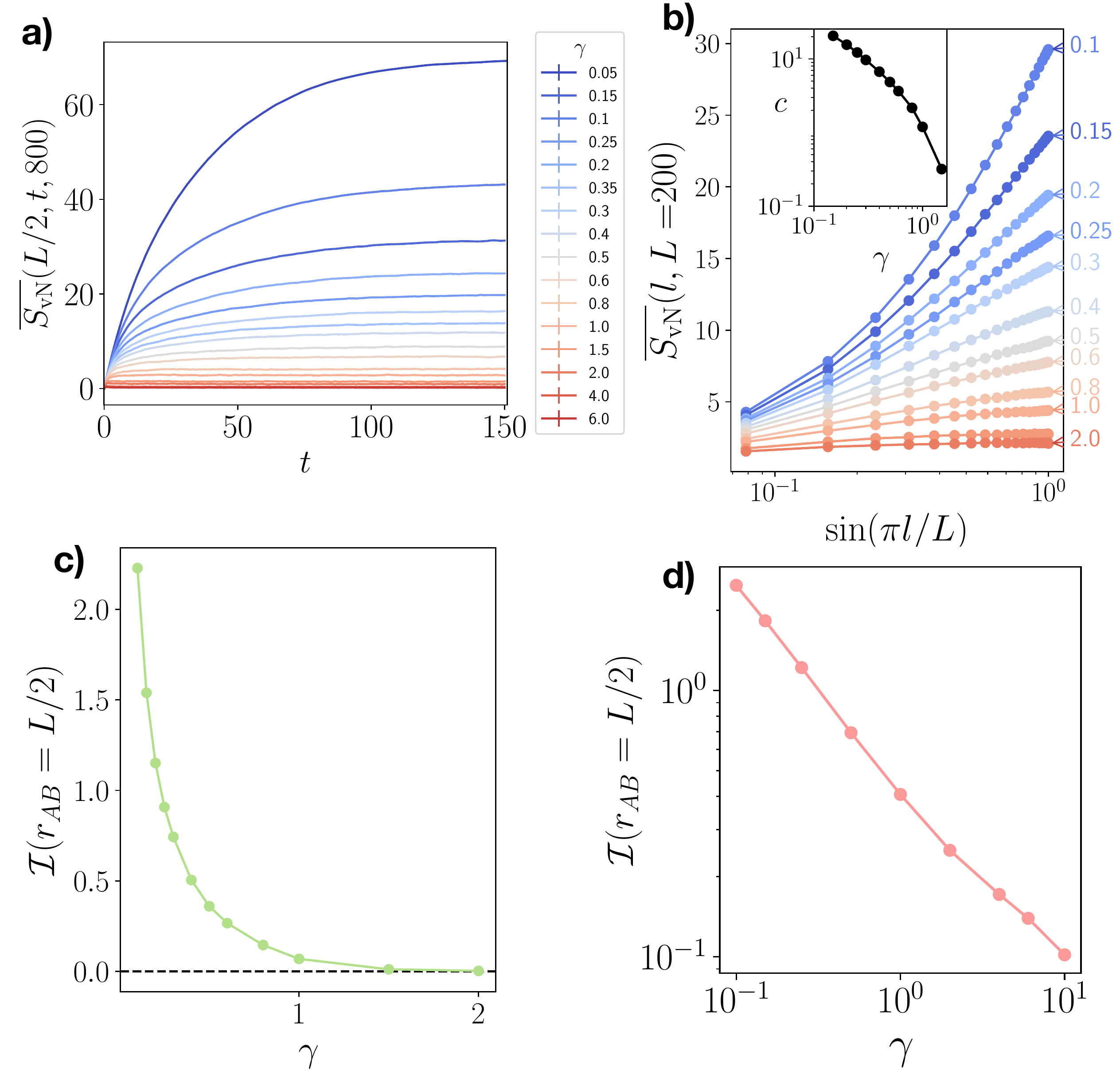}
  \caption{(a) Time-dependence of the trajectory average entanglement entropy for an equal bi-partition of a system of size $L=800$, in the QSD evolution. (b,c) Observables for the QJ evolution (Eq. (2)) for $L=200$. (b) The entanglement entropy as a function of the subsystem size $l$ for different monitoring rates (marked on the figure). The inset shows the effective central charge as a function of $\gamma$. (c) The mutual information of two subsystems with lengths $l_A=l_B=L/8$, with distance $r_{A,B}=L/2$ between their centers. (d) Mutual information of two subsystems with lengths $l_A=l_B=L/8$
    and relative distance $r_{AB}=L/2$ from the QSDc evolution for $L=400$. The mutual information
    $\mathcal{I}$ exhibits a power-law decay with increasing $\gamma$ over the full parameter range and never drops to zero as sharply as in QSD or QJ.} 
  \label{fig:QJ-ent}
\end{figure}

The entanglement entropy,
Fig.~\ref{fig:QJ-ent}(a), and the mutual information, Fig.~\ref{fig:QJ-ent}(b),
in the QJ evolution are qualitatively comparable to the entanglement entropy
obtained from QSD in Fig. 2(a). Both confirm an extended critical regime, where
$\mathcal{I}_{A,B}(r_{AB}=L/2)$ is non-zero and the entanglement entropy grows
logarithmically with system size for weak monitoring ($\gamma\le1$ for $L=200$).
At stronger monitoring, both observables indicate an area law regime where the
mutual information drops to zero. In general we observe that the boundary
between critical and area law behavior seems to be shifted to larger $\gamma$
for the QJ evolution for any system size $L$. However, this contrasts clearly
the absence of an area law phase in the QSDc evolution, as indicated by the
non-vanishing mutual information in Fig.~\ref{fig:QJ-ent} (d).

\section{QSDc evolution, and its connection to non-unitary quantum circuits}

The continuous QSDc dynamics is described by Eq. (1) 
in the main text by setting $\hat M_l=n_l$. It can be seen as a continuous limit of a
non-unitary circuit model, similar to the one considered in
Ref. \cite{chen2020}. Due to the conservation of the total particle number $N$, the dynamics can be simplified to
\begin{equation}
  d| \psi\{\xi_{l,t}\} \rangle=\Big[ -iHdt -\frac{\gamma N dt}{2}+ \sum_l \xi_{l,t} n_l   \Big] |\psi \{\xi_{l,t}\}\rangle.
\end{equation}
The evolution over time-step $d t$ in the limit of small $dt$ can be
approximated via a Trotter decomposition as (apart from a global factor)
\begin{equation}
  \label{eq:non-unitary-circ}
   | \psi(t + dt) \rangle = U_\beta U_{d t} | \psi (t) \rangle  
\end{equation}
with
\begin{align}
 &U_\beta(t) = \exp(- \beta \sum_l \lambda_{l,t} n_l ), \quad \beta = \sqrt{\gamma d t} \nonumber, \\
 &U_{d t}(t) = \exp(- i H dt) ,\nonumber
\end{align}
 where $\lambda_{l,t} \sim \mathcal{N}(0,1)$ is a normally distributed random variable with zero mean and unit variance. This is a random, non-unitary circuit model, which, besides a random Hamiltonian component, is discussed in \cite{chen2020}.

\section{Higher moment evolution}
Observables, which depend on higher moments of the state
$|\psi_t\rangle\langle\psi_t|$ may strongly depend on the specific trajectory
evolution. An example is the entanglement entropy in the main text. Here we
illustrate this with a simple analytical example, the $m$-th moment of the norm
$\langle\psi_t|\psi_t\rangle^m$. We start with the QSD evolution (Eq.~(1) in the main text) and,
for simplicity, a single, Hermitian Lindblad operator $\hat{M}$. The scaling
$\xi_t\sim \sqrt{dt}$ requires that infinitesimal changes are taken into account
up to order $|d \psi_t \rangle^2$. Up to this order, the infinitesimal change is \eq{EqApp1}{
  d\langle\psi_t|\psi_t\rangle^m&=&m\langle\psi_t|\psi_t\rangle^{m-1}(\langle \psi_t|d\psi_t\rangle+\langle d\psi_t|\psi_t\rangle+\langle d\psi_t|d\psi_t\rangle) \nonumber\\
  &&+m(m-1)\langle\psi_t|\psi_t\rangle^{m-2}\left(\langle
    d\psi_t|\psi_t\rangle+\langle\psi_t|d\psi_t\rangle\right)^2. \ \ } The
Hamiltonian evolution cancels out, and expanding again up to order $dt$ one finds
\eq{EqApp2}{
  d\langle\psi_t|\psi_t\rangle^m&=&m\langle\psi_t|\psi_t\rangle^{m-1}\left[(\xi^2-\gamma dt) \langle \psi_t|\hat{M}^2|\psi_t\rangle+\xi\langle \psi_t|\hat{M}|\psi_t\rangle\right] \nonumber\\
  &&+2m(m-1)\langle\psi_t|\psi_t\rangle^{m-2}\xi^2\langle
  \psi_t|\hat{M}|\psi_t\rangle^2. \ \ } The trajectory average thus yields
\eq{EqApp3}{ \overline{d\langle\psi_t|\psi_t\rangle^m}=2\gamma
  m(m-1)\overline{\langle\psi_t|\psi_t\rangle^m\langle\hat{M}\rangle_t^2}dt, }
where $\langle\hat{M}\rangle_t=\langle
\psi_t|\hat{M}|\psi_t\rangle/\langle\psi_t|\psi_t\rangle$. For the first moment,
$m=1$, the term on the right always vanishes, enforcing that the trajectory
averaged norm is constant. Higher moments, however, do generally not vanish and
their evolution depends on the operator $\hat{M}$. For QSDc, $\hat{M}=n$ is the
particle number operator, and one observes in general an exponential growth of
the higher moments with an approximate rate $2\gamma m(m-1)\langle n\rangle^2$.
For QSD, however, $\hat{M}=n-\langle n\rangle_t$ such that $\langle
\hat{M}\rangle_t=0$ for any state and thus any moment $m$ of the norm remains
constant over time.

The norms in the QJ evolution are more involved because here $\xi^2=\xi\sim dt$ and thus arbitrarily high powers in $\xi$ contribute to the evolution of $\langle\psi_t|\psi_t\rangle^m$. We restrict ourselves to $m=1,2$ and again use the operator shortcut $\hat{M}=\left(\frac{n}{\sqrt{\langle n\rangle}}-1\right)$. This yields
\eq{EqApp4}{
d\langle\psi_t|\psi_t\rangle&=&\xi\langle\psi_t|\hat{M}^2+2\hat{M}|\psi_t\rangle=0,\\
d\langle\psi_t|\psi_t\rangle^2&=&\xi\langle\psi_t|\psi_t\rangle\left[\langle\hat{M}\rangle^2+4\langle\hat{M}\rangle\langle\hat{M}^2\rangle+\langle\hat{M}^2\rangle^2\right]=0.
}
Here, only the property $\xi^2=\xi$ was exploited and no trajectory average was required to show that the evolution is constant for this type of jump operator. 

This example can be easily generalized to multiple jump operators $\hat{M}$ and demonstrates that in QSD and QJ trajectories all higher moments of the norm remain constant over time up to order $dt^2$ and an initially normalized state remains normalized. For QSDc on the other hand, higher moments $m>1$ grow roughly exponentially in time, demonstrating that only the average norm of the state is conserved, while its variance is blowing up. 

We emphasize that the difference between the different trajectory evolutions is not just a matter of normalization: the QSDc evolution yields trajectories, which explore a different part of the Hilbert space than the trajectories from QSD and QJ. This difference is not resolved by an \textit{ad hoc} normalization of the state after each numerical time step~\cite{Gisin_1992}.

\section{Auto-correlation functions in the steady-state}
\begin{figure}[h!]
  \centering
  \includegraphics[width=1\linewidth]{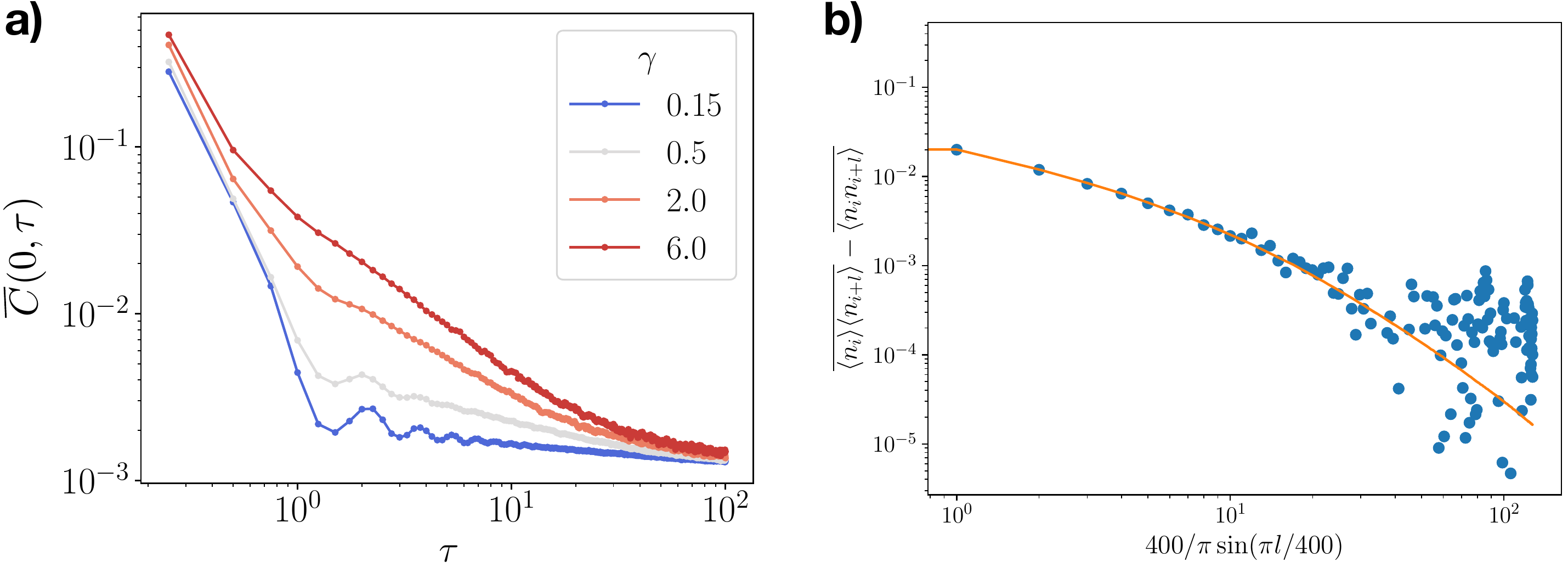}
  \caption{(a) The behavior of the autocorrelation function in the steady-state
    $\overline{C}(0,\tau)$ indicate a quantum-Zeno like evolution with long
    autocorrelation times in the area law regime ($L=400$, QSD trajectories). (b) Visualization of $\overline{C}(l,0)$ from averaging
     the connected density-density correlation function $\overline{\langle n_i n_{i+l}
      \rangle}-\overline{\langle n_i \rangle \langle  n_{i+l}
      \rangle}$ at $L=400, \gamma = 0.35$. This is done by using independent sets of trajectories for
    the two averages $\overline{\langle n_i n_{i+l}
      \rangle}$ and $\overline{\langle n_i \rangle \langle  n_{i+l}
      \rangle}$ (250 trajectories each). The orange line
    shows the value of $\overline{C}(l,0)$ computed from 500 trajectories by direct
    measurement of $|\langle c^\dagger_i c_{i+l} \rangle|^2$.}
  \label{fig:autocorr}
\end{figure}

Further information on the dynamics can be inferred from the autocorrelation function $\overline{C}(0,\tau)$. For unitary, free fermions ($\gamma=0$) it is easy to show that they are given by the Bessel function $\overline{C}(0,\tau)\sim J_0^2(\tau)$, describing damped oscillations with an overall envelope decaying as $\tau^{-1}$. For non-zero $\gamma$ the individual oscillations  in $\overline{C}(0,\tau)$ are more and more suppressed. On the other hand, the overall decay of the auto-correlations slows down, leading to an increased auto-correlation time. When entering the area law regime, the oscillations become over-damped and the auto-correlation time is enhanced significantly, indicating a slowly evolving, quantum-Zeno regime, see Fig.~\ref{fig:autocorr}(a)).

\section{Measurement of $\overline{C}(l,0)$ from the connected density-density correlation function}

In the main text we explored the signatures of the critical regime in $C(l,0)
\equiv | \langle c^\dagger_i c_{i+l} \rangle |^2$ and we discussed that for a Gaussian state this quantity is the connected
density-density correlation function $C(l,0) = \langle n_i \rangle \langle
n_{i+l} \rangle - \langle  n_i n_{i+l} \rangle$. Hence, it is 
possible to obtain the trajectory average of $C(l,0)$ by separately averaging the density-density correlation function
$\overline{\langle n_i n_{i+l} \rangle}$ and product of the
density expectation values $\overline{\langle n_i \rangle \langle  n_{i+l}
  \rangle}$. In Fig.~\ref{fig:autocorr}(b) we show the numerical results of computing $\overline{\langle n_i \rangle \langle  n_{i+l}
  \rangle}$ and $\overline{\langle n_i n_{i+l} \rangle}$ from independent sets
of trajectories. One can see that the data follows the same trend as $\overline{C}(l,0)$ (orange line). The difference of two stochastic variables is more susceptible to fluctuations than the average product  $| \langle c^\dagger_i c_{i+l} \rangle |^2$, which requires a larger number of simulated trajectories.

\section{Entanglement transition in a monitored unitary circuit model}
In order to demonstrate that the entanglement transition reported in the main
text is not restricted to the specific integrable case of free fermions with
uniform nearest-neighbor hopping, we show in this appendix that it extends also
to more generic scenarios. Here we replace the tight-binding Hamiltonian with a
free fermions random-unitary circuit model, which is local in the sense that it
acts only on two neighboring sites, and which preserves the total fermion
number~\cite{chen2020}.

Random-unitary circuit models represent generic unitary dynamics, which mimic
the physics of realistic many-body quantum dynamics, i.e., quantum chaotic
evolution and thermalization, via externally introduced randomness. Indeed the
resulting dynamics is typically highly chaotic and evolves the system towards a
featureless thermal state with volume law entanglement, which is only
constrained by the symmetries of the system.  The random unitary circuit model of noninteracting fermions considered here thus represents a more generic model than the one considered in
the main text as it is non-integrable and contains no quasi-particles and no
other conserved quantities beyond total particle number.

In order to implement a particle number conserving unitary circuit dynamics subject to weak, continuous monitoring of the local particle number, we thus replace the Hamiltonian in the main text with a random circuit counterpart
\eq{EqA1}{
H(t)=\sum_l J_{l,t}(c^\dagger_l c_{l+1}+c^\dagger_{l+1}c_l),
}
where the hopping matrix elements between neighboring sites are now subject to spatiotemporal randomness and are chosen from a binary distribution $J_{l,t}\in \{-1,1\}$ with equal probabilities. We measure time in units of the inverse hopping rate (which is $|J_{l,t}|=1$). The hopping matrix elements are updated after one unit of time in order to maximize the randomness and chaotic evolution on the one hand without trapping the system in a quasi-localized state, which occurs if $J_{l,t}$ changes faster than the particle is hopping. The resulting unitary dynamics are comparable to the one outlined in Ref.~\cite{chen2020}. 

\begin{figure}[h]
 \includegraphics[width=0.99\linewidth]{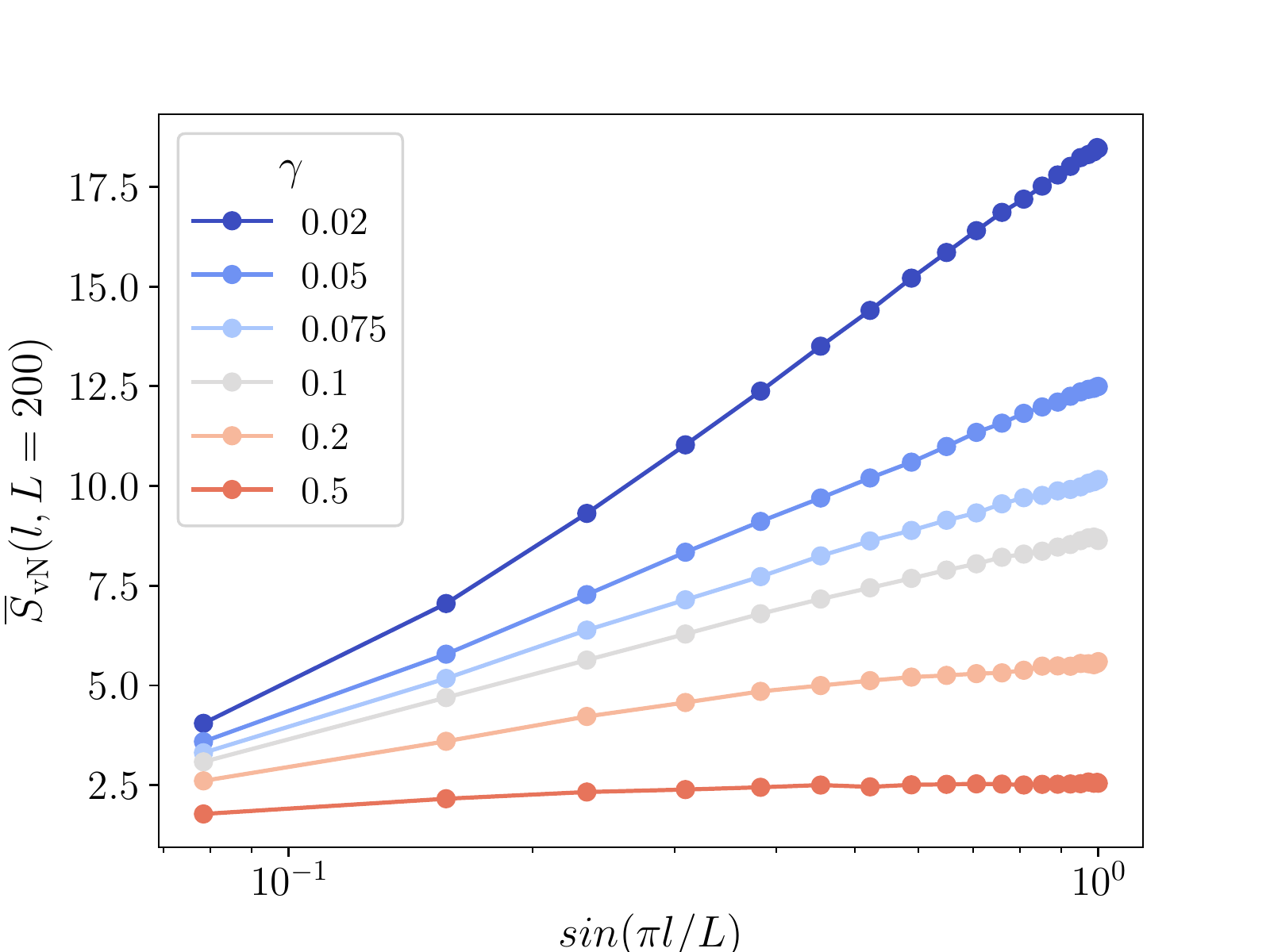} 
  \caption{CFT to area law transition in the random hopping model defined in Eq.~\eqref{EqA1} as a function of the monitoring strength $\gamma$ on a chain of length $L=200$. The entanglement entropy follows a characteristic, logarithmic growth for monitoring rates $\gamma\le 0.2$ and undergoes a transition towards obeying area law for strong monitoring $\gamma>0.2$.}
  \label{fig:randhop}
\end{figure}

The monitoring of the fermion density is implemented via the same QSD protocol we use in the main part of this work, i.e., we only replace $H$ by $H(t)$ in Eq.~(1) in the main text. While the resulting unitary dynamics are clearly non-integrable and does not host well-defined quasi-particles, we observe a behavior of the entanglement entropy in the unitary circuit model which is very similar to the regular tight-binding lattice. Especially, we show in Fig.~\ref{fig:randhop} that the system undergoes an entanglement transition from a CFT-type, logarithmic entanglement growth at low monitoring rates towards an area law phase at large monitoring rates in a system of $L=200$ sites. One can clearly identify a transition in this system. The critical monitoring rate $\gamma_c\approx 0.2$ for this transition is non-universal and depends on the rate with which the unitary disorder acts on the system.

This demonstrates that the CFT to area law transition reported in this paper is
not a peculiar feature of free fermions on a tight-binding lattice with
integrable dynamics, but extends to other, chaotic systems, which share a
similar $U(1)$ symmetry. This claim has just recently been further
substantiated by the analysis of the entanglement entropy dynamics of a
$U(1)$-symmetric, bosonic dynamics \cite{Regemortel2020}, for which the dynamics
is neither free nor integrable.

\section{Trajectory statistics of the entanglement entropy in QSD and QSDc}
\begin{figure}[h]
 \includegraphics[width=0.99\linewidth]{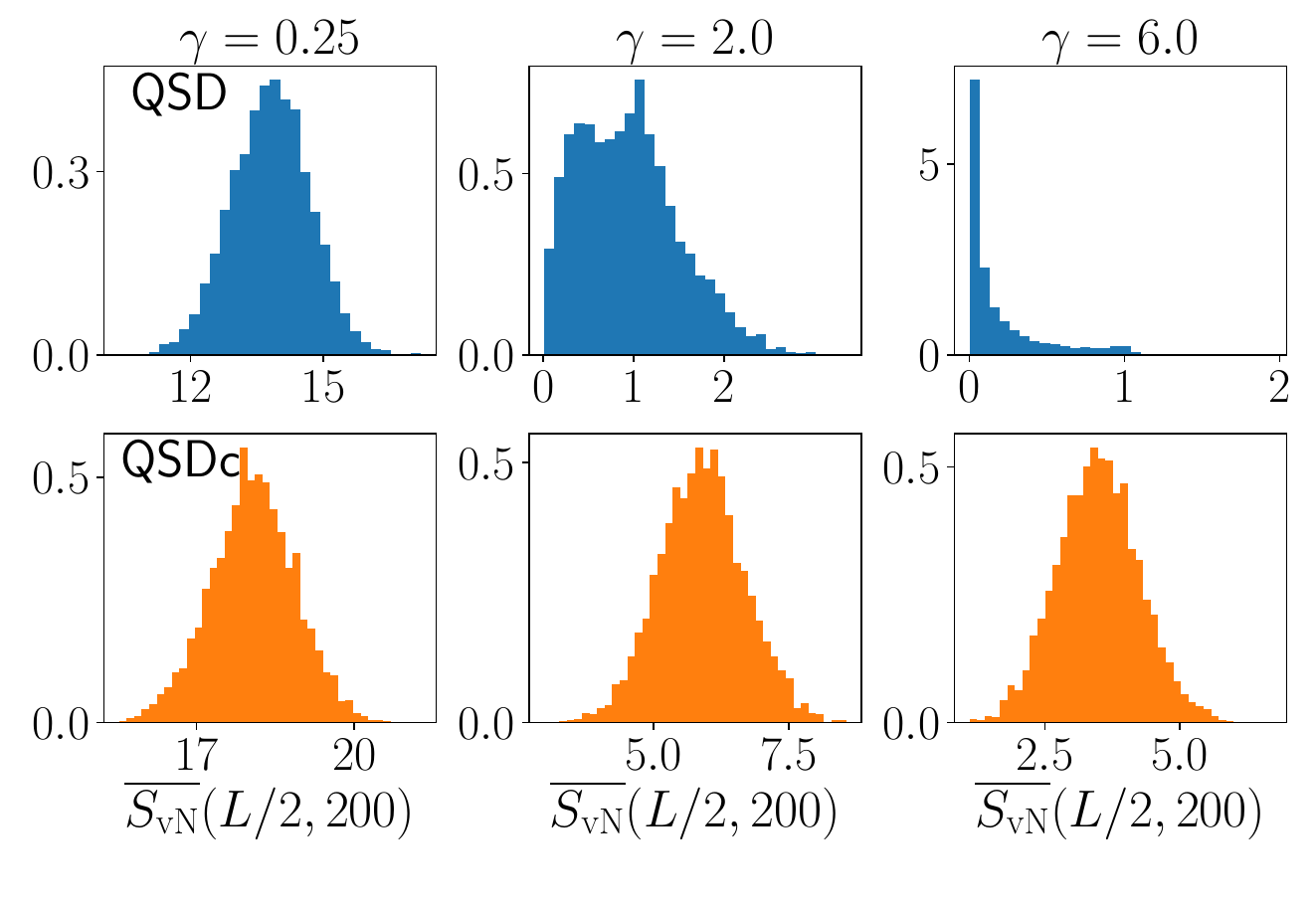} 
  \caption{The trajectory statistics of the entropy reveal a structural difference between the circuit evolution and QSD when the latter is in the area law regime ($L=200, 5000$ trajectories per histogram).}
  \label{fig:statistics}
\end{figure}

We further illustrate the difference between QSD and QSDc by comparing the entanglement entropy distribution for both  evolutions in Fig.~\ref{fig:statistics}. The bins in the histograms reflect the probability for a given entanglement entropy. For weak monitoring, when both types of evolutions predict conformal invariance, both evolutions sample a comparable set of trajectories, i.e., each distribution has similar mean and variance and is symmetric around its peak. The distribution for the QSDc trajectories remains of similar shape for arbitrarily large monitoring rate and only acquires a decreasing mean and variance as $\gamma$ is increased. The distribution of the QSD trajectories, however, undergoes a structural change when it enters the area law phase. It approaches a strongly asymmetric, bimodal distribution with its main peak approaching zero. A second peak emerges and stays pinned at $S_{\text{vn}}=1$, indicating a pronounced probability for finding only a single non-zero eigenvalue of the correlation matrix. This structural difference confirms that QSD and QSDc yield significantly different dynamics for objects with a nonlinear state dependence.

\bibliography{EntEnt}

\end{document}